\documentclass[aps,pra,twocolumn,showpacs,amsfonts,amssymb,amsmath]{revtex4}

\usepackage{graphicx}% Include figure files
\usepackage{dcolumn}% Align table columns on decimal point
\begin{document}
\preprint{draft 1.21}

\title{Upper Limit on the Magnetic Dipole Contribution to the 5$p$-8$p$ Transition in Rb by Use of Ultracold Atom Spectroscopy}

\author{R. Pires}
\author{M. Ascoli}
\author{E. E. Eyler}
\author{P. L. Gould}
\affiliation{Physics Department, University of Connecticut, Storrs, CT 06269, USA}
\author{A. Derevianko}
\affiliation{Physics Department, University of Nevada, Reno, NV, 89557 USA}

\begin{abstract}
We report on hyperfine-resolved spectroscopic measurements of the electric-dipole forbidden 5$p_{3/2} \rightarrow 8p_{1/2}$ transition in a sample of ultracold $^{87}$Rb atoms. The hyperfine selection rules enable the weak magnetic-dipole (M1) contribution to the transition strength to be distinguished from the much stronger electric-quadrupole (E2) contribution. An upper limit on the M1 transition strength is determined that is about 50 times smaller than an earlier experimental determination.  We also calculate the expected value of the M1 matrix element and find that it is less than the upper limit extracted from the experiment.
\end{abstract}
%\submitto{\PRA}
\pacs{32.70.Cs, 31.10.+z, 32.80.Ee}% PACS, the Physics and Astronomy Classification Scheme.
\maketitle

\section{\label{sec:intro}Introduction}
The dominant electronic transitions in atoms are generally electric dipole (E1) transitions, which for single-electron atoms obey the selection rule $\Delta\ell = \pm1$, where $\ell$ is the electron orbital angular momentum. The resonance lines in alkali atoms (e.g., 5$s$$\rightarrow$$5p$ in Rb), with oscillator strengths on the order of unity, are prototypical examples. Obviously, there are many potential transitions that are not allowed by the E1 mechanism. The leading order electric-dipole-forbidden transitions are electric quadrupole (E2) and magnetic dipole (M1) transitions.

Although these dipole-forbidden transitions are generally rather weak, they can in fact be quite useful and important. On the practical side, they expand the number of states that can be accessed by single-photon or multi-photon laser excitation. We have recently exploited this to probe high-lying $d$ states of Rb by single-photon E2 transitions from the $5s$ ground state \cite{Tong2009}. Forbidden transitions can provide narrow resonances and thus play an important role in frequency standards and atomic clocks. In this context, optical-lattice-induced frequency shifts due to such transitions must also be considered \cite{Taichenachev}. States that can decay only by dipole-forbidden transitions are usually metastable and thus play an important role in accumulating population in excited gases such as discharges. On the more fundamental side, these E1-forbidden transitions can be sensitive probes of atomic wavefunctions and thus serve an important role in testing atomic structure calculations. Such tests are important to the interpretation of atomic measurements of fundamental symmetries such as parity violation \cite{Bennett1999,PorBelDer09}. M1 transitions between states of different principal quantum number $n$ are especially useful in this regard as they are nominally subject to the $\Delta n$=0 selection rule. Violations of this rule are due to relativistic and spin-orbit effects. Measuring these M1 transition strengths in various atomic systems and comparing them to theory can help determine the relative importance of contributions such as negative-energy states \cite{Savukov1999}.

In the present work, we explore the E1-forbidden $5p \rightarrow 8p$ transition in Rb by laser excitation of ultracold atoms. The use of ultracold atoms virtually eliminates Doppler broadening and allows hyperfine-selective state preparation \cite{Bhattacharya2003}. With this ability to resolve specific hyperfine transitions, we exploit the associated hyperfine selection rules to separate the E2 and M1 contributions to this E1-forbidden transition. Specifically, we use $^{87}$Rb, with a nuclear spin $I$=3/2, and prepare atoms in the $5p_{3/2}(F'$=0) state. As shown in Fig. 1, and derived in Sec. \ref{sec:theory}, the transition to $8p_{1/2}(F''$=1) is allowed only by the M1 process, while the transition to $8p_{1/2}(F''$=2) is only E2 allowed.

Our work was motivated in part by the surprising results of Bayram, \textit{et al.} \cite{Bayram2000}. They used a polarization technique to measure the relative E2 and M1 contributions to the Rb $5p \rightarrow 8p$ transition and obtained an anomalously large M1 fraction, at odds with theoretical calculations. Here we resolve this discrepancy in favor of theory, setting an upper limit on the M1 contribution which is larger than, and therefore consistent with, the predicted value.

There have been a number of other E2/M1 ratio measurements in various atomic systems. For example, in atomic bismuth, fits to the hyperfine spectra of light emitted by a discharge revealed the relative importance of E2 and M1 transitions \cite{Heldt1968,Werbowy2007}. The M1-E2 interference effect in the Zeeman spectra has been utilized in atomic lead \cite{Jenkins1941,Werbowy2009}. In atomic thallium, measurements of absorption spectra and Faraday rotation, combined with lineshape modeling, yielded the E2/M1 ratio \cite{Majumder1999}. In all of these examples, the M1 process was dominant because the transitions took place within the same ground-state electronic configuration. In our case, both initial and final states are electronically excited and the transition changes $n$, so the M1 process is significantly suppressed. Our work also differs in that the E2-only and M1-only transitions are well resolved in our excitation spectra, obviating the need for sophisticated modeling. We note that there have also been proposals for measuring E2/M1 ratios in alkali $n_1p-n_2p$ transitions using polarization effects in two-wave mixing \cite{Mironova2005a,Mironova2005b}.

This paper is organized as follows. In Sec. \ref{sec:theory}, we derive the E2 and M1 hyperfine selection rules, describe the calculations of the E2 and M1 matrix elements, and predict the signal ratio for relevant hyperfine transitions. The experimental setup is described in Sec. \ref{sec:experiment}, and the experimental results and analysis in Sec. \ref{sec:results}. Section \ref{sec:conclusions} comprises concluding remarks.

\section{\label{sec:theory}Theory}

The Hamiltonians for magnetic dipole and electric quadrupole transitions may be written in atomic units (a.u.) as \cite{Johnson2007,Weissbluth1978}
\begin{eqnarray}
H_\textrm{M1}  &=& \textbf{M}\cdot\textbf{B},\textrm{ and} \\
H_\textrm{E2}  &=& \textbf{E} \cdot \hat{Q} \cdot \textbf{k},
\end{eqnarray}
where \textbf{E} and \textbf{B} are the electric and magnetic field amplitudes, respectively, and \textbf{k} is the wavevector. In Eq. (1), $\textbf{M}=\frac{1}{2}(\textbf{L}+g_S\textbf{S})$  is the non-relativistic magnetic dipole operator, where \textbf{L} and \textbf{S} are the spin and orbital angular momentum, respectively, and $g_s=2.00232$ is the electron $g$-factor. Notice that the expressions for the magnetic moments differ in the Gaussian and the SI systems of electromagnetic units by a factor of the speed of light. In the conversion to atomic units this translates into a factor of $\alpha=1/137$. Below we use the SI system of electromagnetic units, in which the Bohr magneton is $\mu _B = e\hbar/(2m_e)$, or 1/2 atomic unit. The operator $\hat{Q} = e\,r^2C^{(2)}$ in Eq.(2) is the electric quadrupole operator, where $C^{(2)}$  is a spherical harmonic tensor and $r$ is the radial coordinate of the electron.

We focus here on the situation relevant to our experiment, transitions betweens specific hyperfine levels $F$ and magnetic sublevels $m$: $5p_{3/2}(F',m') \rightarrow 8p_{1/2}(F'',m'')$. The hyperfine selection rules for M1 transitions can be derived by applying the Wigner-Eckart theorem to the corresponding matrix elements for a spherical component $q$ of the magnetic dipole operator \textbf{M}, then uncoupling the nuclear spin angular momentum from the electronic angular momentum:
\begin{widetext}
\begin{equation}
  \left\langle {8p_{1/2} F''m''} \right|{M_q}\left| {5p_{3/2} F'm'} \right\rangle  = a \left(\hspace{-4pt} {\begin{array}{c c c}
   {F''} & 1 & {F'}  \\
   { - m''} & q & {m'}  \\
 \end{array} } \hspace{-3pt}\right)(2F' + 1)(2F'' + 1)
   \left\{\hspace{-4pt} {\begin{array}{c c c}
   {J'} & {F'} & I  \\
   {F''} & {J''} & 1  \\
 \end{array} } \hspace{-3pt}\right\}\left\langle {8p_{1/2} } || \textbf{M} || {5p_{3/2} } \right\rangle .
\end{equation}
\end{widetext}
Here $a$ is a phase factor, $a=(-1)^{F''-m''}$, the term in parentheses is a Wigner 3-$j$ symbol, and the term in curly brackets is a 6-$j$ symbol. The selection rules
\begin{eqnarray}
\Delta F &=& 0, \pm 1,\\
\Delta m &=& 0, \pm 1,
\end{eqnarray}
and the triangle rule
\begin{equation}
F' + F'' \geq 1
\end{equation}
follow from the 3-$j$ symbol and the vector nature of \textbf{M}, which constrains $q$ to the values $q=0,\pm1$. Repeating this procedure for the case of E2 transitions yields a similar expression for the matrix element of the $q$'th component of the spherical harmonic tensor $C_q^{(2)}$ in the electric quadrupole operator $\hat{Q}$,

\begin{widetext}
\begin{equation}
  \left\langle {8p_{1/2} F''m''} \right|{Q_q}\left| {5p_{3/2} F'm'} \right\rangle  = a
  \left(\hspace{-4pt} {\begin{array}{c c c}
   {F''} & 2 & {F'}  \\
   { - m''} & q & {m'}  \\
 \end{array} } \hspace{-3pt}\right)(2F' + 1)(2F'' + 1)
   \left\{\hspace{-4pt} {\begin{array}{c c c}
   {J'} & {F'} & I  \\
   {F''} & {J''} & 2  \\
 \end{array} } \hspace{-3pt}\right\}\left\langle {8p_{1/2} } || \hat{Q} || {5p_{3/2} } \right\rangle.
\end{equation}
\end{widetext}
This result yields the E2 selection rules
\begin{eqnarray}
\Delta F &=& 0, \pm 1, \pm 2,\\
\Delta m &=& 0, \pm 1, \pm 2,
\end{eqnarray}
and the triangle rule
\begin{equation}
F' + F'' \geq 2.
\end{equation}
If the electric field and the initial atomic state are polarized, there are additional constraints on $\Delta m$ due to the double dot-product in Eq. (2).  This can also constrain the allowed values of $F''$ under some circumstances.

Thus, starting in an $F'$=0 state, an M1 transition to $F''$=2 is forbidden because of Eq. (4), while an E2 transition to $F''$=1 is forbidden because of Eq. (10). We thereby obtain the key result, shown in Fig. 1, that $5p_{3/2}(F'$=0$) \rightarrow 8p_{1/2}(F''$=1) is solely an M1 transition, while $5p_{3/2}(F'$=0$) \rightarrow 8p_{1/2}(F''$=2) is only E2 allowed. The various allowed transitions for $5p_{3/2}(F') \rightarrow 8p_{1/2}(F'')$  are shown more generally in Table I for an unpolarized sample.
\begin{figure}
\centering \vskip 0 mm
\includegraphics[width=0.9\linewidth]{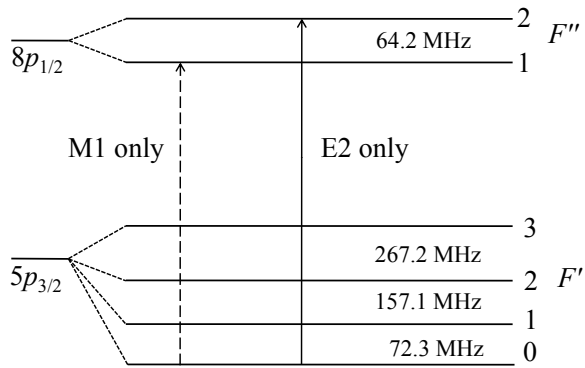}
\caption{\protect\label{fig1} Hyperfine transitions used to separate the M1 and E2 contributions to the $^{87}$Rb 5$p_{3/2}(F') \rightarrow 8p_{1/2}(F'')$ transition at 587.6 nm. Starting in $F'$=0, the transitions to $F''$=1 (dashed line) and $F''$=2 (solid line) are M1 allowed and E2 allowed, respectively. The indicated hyperfine splittings are not drawn to scale}
\end{figure}
\begin{table}
\caption{Allowed $5p_{3/2}(F')\rightarrow 8p_{1/2}(F'')$ transitions in $^{87}$Rb. Note that our use of $F'$=0 allows complete separation of M1 and E2 processes as shown in the first row (boldface).}

\begin{ruledtabular}
\begin{tabular}{c c c}

 & $F''$=1 & $F''$=2 \\
 \hline
$F'$=0  &  \textbf{M1} & \textbf{E2} \\
$F'$=1  &  M1+E2 & M1+E2 \\
$F'$=2  &  M1+E2 & M1+E2 \\
$F'$=3  &  E2 & M1+E2 \\
 \end{tabular}
\end{ruledtabular}
\end{table}

In the experiment, we measure signals for specific $F'\rightarrow F''$ transitions. If the initial 5$p$ state is prepared without orientation or alignment, which should be a good approximation under our conditions as we discuss in Section IV, the transition rates for M1 and E2 processes will be proportional to the Einstein $B$ coefficients.  They can be obtained \cite{Johnson2007} by summing over $m''$ for a given $m'$ and then averaging over $m'$:
\begin{equation}
B_{\textrm{M1}}  = \frac{4\pi^2}{3c^2} \sum\limits_{m',m''}
{\frac{{\left| {\left\langle {8p_{1/2} F''m''} \right|\textbf{M}\left|
{5p_{3/2} F'm'} \right\rangle } \right|^2 }} {2F' + 1}}
\end{equation}
\begin{equation}
B_{\textrm{E2}} = \frac{\pi^2k^2}{15} \sum\limits_{m',m''}
{\frac{{\left| {\left\langle {8p_{1/2} F''m''} \right|\hat{Q}\left|
{5p_{3/2} F'm'} \right\rangle } \right|^2 }} {2F' + 1}}.
\end{equation}
For the special case of an $F'$=0 initial state, used for most of the results reported here, the averaging is redundant as there is only one magnetic sublevel and the state cannot support either orientation or alignment.  In this case the M1 (E2) transition strength is associated only with the transition to $F''$=1 ($F''$=2) and the above expressions can be simplified to
\begin{eqnarray}
B_{\textrm{M1}}&=& \frac{\pi^2}{9c^2} \left| {\left\langle {8p_{1/2} } || \textbf{M} || {5p_{3/2} } \right\rangle } \right|^2
\\
B_{\textrm{E2}} &=& \frac{\pi^2k^2}{300} \left| {\left\langle {8p_{1/2}} || \hat{Q} || {5p_{3/2} } \right\rangle } \right|^2.
\end{eqnarray}
The ratio $R=B_{\textrm{M1}}/B_{\textrm{E2}}$ is measured in the experiment by determining the ratio of signal sizes at a fixed laser intensity.  Assuming that the E2 electric quadrupole moment matrix element is known, we can determine the M1 magnetic dipole moment matrix element.

	Non-relativistically, the M1 matrix element between the $5p$ and $8p$ states vanishes identically, because the non-relativistic M1 operator does not couple the radial motion of the electrons and the non-relativistic radial wavefunctions of the $5p$ and $8p$ states are orthogonal ($\Delta n$=0 selection rule). In other words, the entire value of the M1 $5p_{3/2}\rightarrow 8p_{1/2}$ matrix element is due to relativistic effects. Qualitatively, this may be understood from the fact that the radial wavefunctions for the $p_{1/2}$ and $p_{3/2}$ angular symmetries arise from solving separate radial Dirac equations and the resulting $5p_{3/2}$ and $8p_{1/2}$ radial orbitals are no longer orthogonal. A general relativistic treatment of multipolar transitions may be found, for example, in Ref. \cite{Johnson2007}. Unfortunately, there are errors in the relevant formulae of that book and here we present revised expressions for the single-particle reduced matrix elements for the M1 and E2 operators:
\begin{eqnarray}
\langle n''\kappa''||\textbf{M}||n'\kappa'\rangle = \frac{\kappa''+\kappa'}{2\alpha}\,
\langle-\kappa''||C^{\left(1\right)}||\kappa'\rangle\times \nonumber\\*
\int_{0}^{\infty}\left(  G_{n''\kappa''}\left(  r\right)
F_{n'\kappa'}\left(  r\right)  +F_{n''\kappa''}\left(  r\right)
G_{n'\kappa'}\left(  r\right)  \right)  rdr,
\end{eqnarray}
\begin{eqnarray}
 \langle n''\kappa''||\hat{Q}||n'\kappa'\rangle = \langle\kappa''||C^{\left(2\right)}
||\kappa'\rangle\times \nonumber\\*
 \int_{0}^{\infty}\left(  G_{n''\kappa''}\left(  r\right)  G_{n'\kappa'}\left(
r\right)  +F_{n''\kappa''}\left(  r\right)  F_{n'\kappa'}\left(
r\right)  \right) r^{2}dr.
\end{eqnarray}
Here $C^{(k)}$ are the normalized spherical harmonics, $G(F)$ are the large
(small) radial components of  the Dirac bi-spinor, and the angular quantum
number $\kappa$ is \ defined as  $\kappa=(l-j)(2j+1)$. Both expressions are
given in the long-wavelength approximation (transition wavelength is much
larger than the atomic size). In addition, the E2 operator is given in the
conventional length gauge, which is more stable in calculations and does not
require introducing so-called derivative terms into the many-body calculations.

We evaluate the matrix elements using two methods: the Dirac-Hartree-Fock (DHF) approximation and a more elaborate Singles-Doubles (SD) method \cite{SafJohDer99}. Both calculations are \textit{ab initio} relativistic. The SD method builds upon an expansion of the atomic many-body wavefunction into a large series of the Slater determinants classified by the number of excited electrons (two at most for the SD) from the reference DHF state. Major correlation corrections (e.g., core polarization) are included. Previously, the SD method has been shown \cite{SafJohDer99} to yield theoretical E1 matrix elements for principal transitions in Rb with a sub-1\% accuracy. While more sophisticated methods are available \cite{PorBelDer09}, this accuracy is sufficient for the goals of this paper.  Numerical evaluation has been carried out using the dual-kinetic-balance basis set \cite{BelDer08} generated in a large, 100 a.u., cavity. SD runs employed 66 out of 75 basis functions for each partial wave up to $\ell$ = 5. We find that the DHF removal energies are off by 5\% for the $5p_{3/2}$ and 2\% for the $8p_{1/2}$ states. Further application of the SD method (i.e., accounting for correlations) brings the theoretical energies into a 0.5\% and 0.1\% agreement with experimental values. Finally, the M1 and E2 reduced matrix elements are listed in Table II. We find that these matrix elements are largely insensitive to correlations (the M1 (E2) matrix element is affected at the 11\% (8\%) level). Considering that the SD method accounts for the dominant many-body effects, this indicates that the theoretical accuracy of the SD matrix elements is at the level of 1\% or better.
\begin{table}
\caption{Values for M1 and E2 reduced matrix elements, in atomic units.}

\begin{ruledtabular}
\begin{tabular}{c c c}

 & $\left| {\left\langle {5p_{3/2} } || \textbf{M} || {8p_{1/2} } \right\rangle } \right|$ &
 $\left| {\left\langle {5p_{3/2} } || \hat{Q} || {8p_{1/2} } \right\rangle } \right|$\\
 \hline
DHF  &  $1.95 \times 10^{-3}$ & 7.05 \\
SD  &  $2.16 \times 10^{-3}$ & 6.53 \\
 \end{tabular}
\end{ruledtabular}
\end{table}

\section{\label{sec:experiment}Experiment}
The experiment is performed in a diode-laser-based vapor-cell magneto-optical trap (MOT), which is used to trap $\sim$$10^7$ $^{87}$Rb atoms at a temperature of $\sim$100~$\mu$K with a density up to $10^{11}$ cm$^{-3}$ \cite{Tong2009}. As in a standard MOT, the trapping light is tuned 13 MHz below the $5s_{1/2}(F$=2$)\rightarrow 5p_{3/2}(F'$=3) cycling transition (see Fig. 2(a)). However, the repumping light, which prevents accumulation of population in $5s_{1/2}(F$=1), is tuned midway between the $5s_{1/2}(F$=1$)\rightarrow 5p_{3/2}(F'$=1) and $5s_{1/2}(F$=1$)\rightarrow 5p_{3/2}(F'$=2) transitions. This allows us to lock this laser to the corresponding crossover resonance in the saturated absorption spectrum. More importantly, it enables the subsequent population transfer described below to be done conveniently with readily available acousto-optical modulators (AOMs). This sub-optimal MOT configuration results in only $\sim$20\% fewer atoms than the standard tuning of the repumping laser to exact resonance with $5s_{1/2}(F$=1$)\rightarrow 5p_{3/2}(F'$=1 or 2).

To spectroscopically separate the M1 and E2 contributions to the $5p_{3/2}\rightarrow 8p_{1/2}$ transition, the atoms must be prepared in the $F'$=0 state of $5p_{3/2}$. This is done with the excitation steps shown in Fig. 2, with the timing sequence shown in Fig. 3. First, the trapping and repumping beams are switched off with AOMs, leaving the atoms in $5s_{1/2}(F$=2). Next, a depletion beam tuned to $5s_{1/2}(F$=2$)\rightarrow 5p_{3/2}(F'$=2) is switched on in order to optically pump the atoms into $5s_{1/2}(F$=1). This beam is derived from the trapping laser by frequency shifting with an AOM. Finally, a preparation beam, derived from the repumping laser, is switched on to excite the $5s_{1/2}(F$=1$)\rightarrow 5p_{3/2}(F'$=0) transition. The $5p_{3/2}\rightarrow 8p_{1/2}$ transition is excited with 587.6 nm light from a cw single-frequency tunable ring dye laser system (Coherent 699-29 with Rhodamine 6G dye) pumped by an argon-ion laser. This light is left on throughout the excitation cycle.
\begin{figure}
\centering \vskip 0 mm
\includegraphics[width=0.95\linewidth]{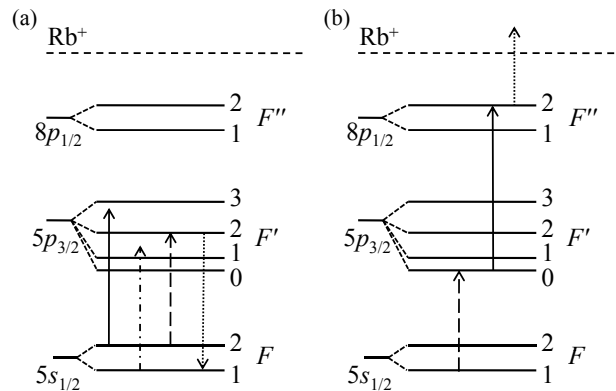}
\caption{\protect\label{fig2} Level diagrams showing the transitions used for state preparation and $5p_{3/2}(F')-8p_{1/2}(F'')$ spectroscopy. (a) The trap (solid) and repump (dashed-dotted) lasers provide an ultracold sample of $^{87}$Rb atoms which are optically pumped into $5s_{1/2}(F$=1) using the depletion laser (dashed) and spontaneous emission (dotted). (b) The preparation laser (either pulsed or cw, dashed) provides excitation to $5p_{3/2}(F'$=0), and the tunable dye laser at 587.6 nm (solid) further excites to $8p_{1/2}(F''$=1,2). Atoms in $8p_{1/2}$ are detected by pulsed photoionization at 1064 nm (dotted).}
\end{figure}
\begin{figure}
\centering \vskip 0 mm
\includegraphics[width=0.95\linewidth]{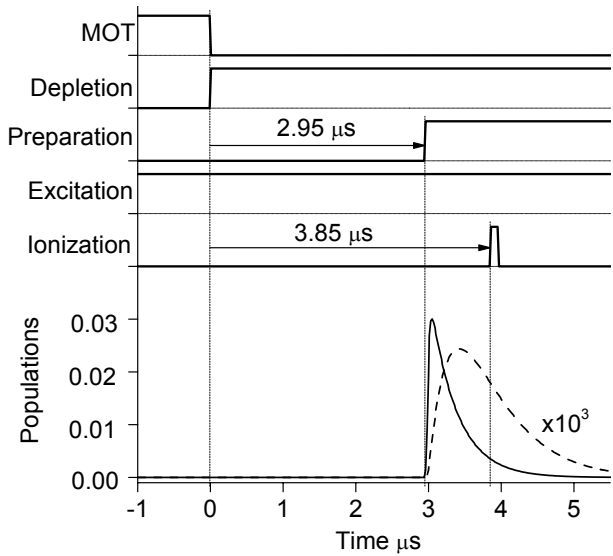}
\caption{\protect\label{fig3} Timing diagram for the $5p_{3/2}(F')-8p_{1/2}(F''$) spectroscopy. The case of pulsed preparation ($5s_{1/2}-5p_{3/2}$ excitation) is shown. Populations in the $5p_{3/2}(F'$=0) (solid line) and $8p_{1/2}(F''$=2) (dashed line) levels, calculated by rate equations, are also shown. The preparation and excitation intensities are 5.2 mW/cm$^2$ and 62 W/cm$^2$, respectively.}
\end{figure}

Although the $5s_{1/2}(F$=1$)\rightarrow 5p_{3/2}(F'$=0) transition is a cycling transition with respect to $F$ (i.e., $F'$=0 can only decay back to $F$=1), the $F'$=0 population is expected to be transient due to optical pumping. Because the $5s_{1/2}(F$=1) level has a higher degeneracy than the $5p_{3/2}(F'$=0) level, after a few cycles of excitation and spontaneous decay, most atoms are pumped into dark magnetic sub-levels and cannot be further excited. If the excitation light is $\pi$-polarized and drives $\Delta m$=0 transitions, the population accumulates in $5s_{1/2}(F$=1, $m$=$\pm$1). In order to determine the optimum timing for the various beams in the presence of this optical pumping, we solve the rate equations for the time-dependent level populations of the system. The $5s_{1/2}(F$=1,2), $5p_{3/2}(F'$=0,1,2,3), and $8p_{1/2}(F''$=2) levels are included. In addition, $5s_{1/2}(F$=1, $m$=$\pm$1) is included as a dark state that can be populated by spontaneous decay, but not excited. Appropriate lifetimes, transition strengths and branching ratios are used in the simulation, as well as the experimental values for intensities and timings. The long-lived 8$p_{1/2}$ level (480 ns lifetime \cite{Theodosiou84}) is assumed to decay back to $5p_{3/2}$ via $6s_{1/2}$.  The resulting $5p_{3/2}(F'$=0) and $8p_{1/2}(F'$=2) populations, shown in Fig. 3, reveal an optimum preparation time before optical pumping into the dark state occurs. However, as discussed in Sec. \ref{sec:results}, this dark-state pumping is not as severe in the experiment as these calculations predict.

For detection of $8p_{1/2}$ atoms, we use photoionization with an injection-seeded pulsed Nd:YAG at 1064 nm. The fluence of each 5~ns FWHM pulse is typically 0.11~J/cm$^2$, yielding an estimated $8p_{1/2}$ ionization probability of 63\%. The entire excitation sequence is synchronized with the 10~Hz repetition rate of this pulsed detection laser. We note that the 532 nm second harmonic on the Nd:YAG laser was originally used for $8p_{1/2}$ ionization, but this yielded a significant background due to two-photon $5s_{1/2}$ ionization. Using the 1064 nm light mitigates this problem, since $5s_{1/2}$ ionization now requires four photons. Also, lower pulse energies can be used because the 1064 nm photoionization cross section, $1.62 \times 10^{-18}$ cm$^2$ (calculated using phase-shifted Coulomb wave functions), is a factor of 6.4 larger than that at 532 nm.

The MOT is located between a pair of 95\% transparent grids separated by 2.09 cm. A pulsed electric field of $\simeq$216~V/cm is applied about 1.1~$\mu$s after the photoionization pulse in order to extract the resulting photoions. These ions are detected with a discrete dynode electron multiplier (ETP model 14150) whose output is sent to a boxcar averager in order to select the desired time-of-flight window. A computer records the data from the averager as the $5p_{3/2}\rightarrow 8p_{1/2}$ laser is scanned.

\section{\label{sec:results}Results}

A typical spectrum for the $5p_{3/2}(F'$=0$)\rightarrow 8p_{1/2}$\linebreak[1]$(F''$=1,2) transitions is shown in Fig. 4(a). The $5p_{3/2}(F'$=0$)\rightarrow 8p_{1/2}(F''$=2) E2-allowed transition is clearly seen, while the $5p_{3/2}(F'$=0$)\rightarrow 8p_{1/2}(F''$=1) M1-allowed transition is not observable above the noise. This allows us to set a limit on the M1 matrix element. For comparison, we show in Fig. 4(b) a scan over the $5p_{3/2}(F'$=2$)\rightarrow 8p_{1/2}(F''$=1,2) transitions, both of which are E2 allowed. Since the depletion light transiently populates $5p_{3/2}(F'$=2), and the 8$p_{1/2}$ excitation light is always on, detected ions are associated with transitions from this level. This scan is useful on two accounts. First, it allows us to check the expected ratio of $m$-averaged transition strengths for these two E2-allowed transitions. We measure a ratio of areas under these peaks of 0.43$\pm$0.04, which compares well with the predicted value of 0.43.  This prediction is based on a homogeneous distribution of population in the magnetic sublevels, a reasonable assumption in the complex radiative environment of the MOT. Second, it provides an accurate calibration of the 64.24~MHz $8p_{1/2}(F''$=1,2) splitting \cite{Tsekeris1975}, thereby telling us where in Fig. 4(a) the $5p_{3/2}(F'$=0$)\rightarrow 8p_{1/2}(F''$=1) M1-allowed transition should be located.

Due to nonlinearity in the scanning of the dye laser, an uncertainty in this frequency calibration on the order of a few percent is expected. We verify that this uncertainty is $<$5\% by comparing scans over the $5p_{3/2}(F'$=2$)\rightarrow 8p_{1/2}(F''$=1,2) and $5p_{3/2}(F'$=3$)\rightarrow 8p_{1/2}(F''$=1,2) transitions. The latter is obtained by leaving the MOT beams on and keeping the depletion and preparation beams off, which results in populating the $5p_{3/2}(F'$=3) level. The uncertainty in the $8p_{1/2}(F''$=1,2) splitting due to scan nonlinearity is taken into account in the fits that are used to set a limit on the M1 transition.
\begin{figure}
\centering \vskip 0 mm
\includegraphics[width=0.8\linewidth]{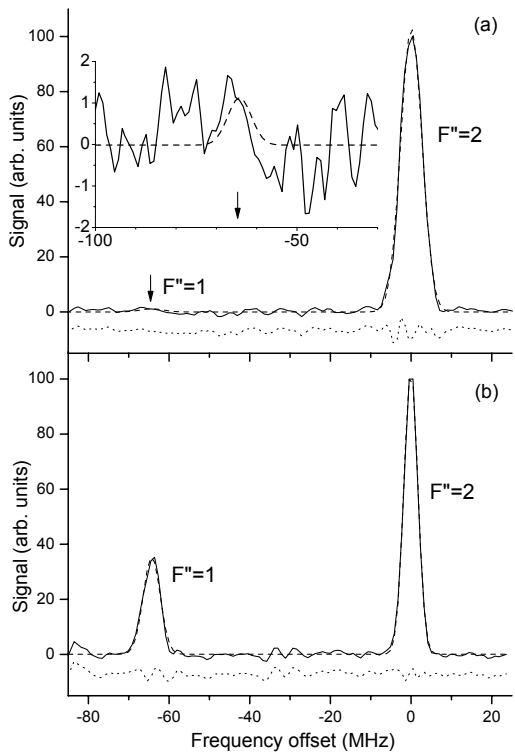}
\caption{\protect\label{fig4} (a) Solid line: Typical spectrum of the $5p_{3/2}(F'$=0) $\rightarrow$ $8p_{1/2}(F''$=1,2) transition. As shown in the inset, no signal is observed for $F''$=1 (near -64 MHz), setting a limit on the relative M1 transition strength. Dashed line: Best fit to a pair of Gaussians, as described in text.  Dotted line: residuals from the fit, offset by seven units for visibility.  (b) Spectrum of the $5p_{3/2}(F'$=2$)\rightarrow 8p_{1/2}(F''$=1,2) transition (solid line), with best fit to a pair of Gaussians (dashed line), and residuals offset by seven units (dotted line). In this case, allowed E2 transitions to both $F''$=1 and 2 are observed.}
\end{figure}

As discussed in Sec. \ref{sec:experiment}, we expect that optical pumping into the $5s_{1/2}(F$=1) dark state will result in an optimal timing of the $5s_{1/2}(F$=1$)\rightarrow 5p_{3/2}(F'$=0) preparation pulse. However, we find that a cw preparation beam actually yields a larger signal (by a factor of $\sim$2) and a better signal-to-noise than a pulsed preparation beam. We believe that this counter-intuitive suppression of the optical pumping is likely caused by multiple scattering of resonance fluorescence light in the optically dense sample and/or Zeeman precession of the spin in the quadrupole field of the MOT \cite{Balik2009}.

To set a limit on the strength of the M1 transition, we fit the observed $5p_{3/2}(F'$=0$)\rightarrow 8p_{1/2}(F''$=2) E2 signal and the unobserved $5p_{3/2}(F'$=0$)\rightarrow 8p_{1/2}(F''$=2) M1 signal to a pair of Gaussians separated by the 64 MHz $8p_{1/2}(F''$=1,2) splitting. Although we do not understand the lineshape of the observed transition in detail, the scans fit quite well to Gaussians with a typical full width at half-maximum (FWHM) of 10 MHz, as seen in Fig. 4(a). The 480 ns \cite{Theodosiou84} lifetime of the $8p_{1/2}$ upper state of this two-photon transition would yield a Lorentzian natural linewidth of 330 kHz. However, the two lasers driving this transition each have linewidths in the range of 1-2 MHz. Also, the inhomogeneous magnetic field sampled by the atoms in the MOT results in a distribution of Zeeman shifts for the various magnetic sublevels. The fact that the first step of the two-photon transition is on resonance and close to saturation may also affect the lineshape. The second step (E2) transition is very weak, with a transition rate estimated to be $\sim$10$^4$ times smaller than that of the first step (E1) transition. The fits are clearly not consistent with a Lorentzian lineshape with the 5.9 MHz FWHM natural linewidth of the $5p_{3/2}$ intermediate state. Since the observed peaks for other transitions, such as those shown in Fig. 4(b), are also Gaussian with similar widths, we feel justified in using a Gaussian fit for setting a limit on the unobserved M1 transition. Understanding the details of the lineshape and its dependence on various parameters is an interesting topic for further investigation.

An upper limit on the M1 transition strength is determined by least-squares fitting of a number of individual scans to a pair of Gaussians. The separation between the peaks is constrained to $64\pm3$ MHz, and the FWHM of each Gaussian is allowed to vary independently from 7 to 21 MHz. We also include constant and linear terms in the fits to account for the background and its slow drift, respectively. A total of 32 scans are included, 20 of them using pulsed preparation and the remainder using cw preparation. An example of the latter is shown in Fig. 4(a). For each scan we determine the ratio $R$ of amplitudes for the two Gaussians, for comparison with the theoretical prediction obtained using Eqs. (13-14) and the matrix elements in Table II.  The mean experimental ratio is found to be $R$=0.0026 and the standard deviation of the mean is 0.0020. Assuming a normal distribution of values, we set an upper limit on $R$ by adding to the mean two standard deviations of the mean. This yields $R<$0.007 with 95\% confidence. By this criterion, our result is consistent with zero. We verify that omitting the linear background term from the fits has a minimal effect, changing the mean value of $R$ to 0.0024 and its upper limit to 0.006. We perform a similar statistical analysis of the scans by simply integrating the signal in 6.4 MHz wide bins instead of doing Gaussian fits. This yields a mean value of $R$=0.0021 and an upper limit of 0.005. As our final value, we quote $R<$0.007.

Statistics connected to an F-test \cite{Freund1992} reveal that the Gaussians fitted to the position of the M1 peak are not significant, i.e., the probability of an actual peak at the expected location is well below 95\%. When we add a simulated peak to our data, the F-test shows that, with our present noise level, we need to add a signal with $R>$0.01 in order for it to be significant.

In Table III we summarize our results.  In our experiment, the limiting value of the ratio $R$ of signals for the $5p_{3/2}(F'$=0$)\rightarrow 8p_{1/2}(F''$=2) M1-allowed transition and the $5p_{3/2}(F'$=0$)\rightarrow 8p_{1/2}$\linebreak[1]$(F''$=2) E2-allowed transition is determined directly. The previous polarization-based measurements \cite{Bayram2000} can be converted to yield an indirect value for $R$, shown in the middle column of Table III. Obviously, our upper limit on $R$ is well below the range reported in this earlier work. Also shown is our theoretical prediction.  The experimental limit is consistent with this theoretical prediction, but we are not able to provide a statistically significant non-zero result with which to compare the theory. Table III also compares the M1 matrix elements resulting from our measurement with the earlier experiment and the theory.
\begin{table}
\caption{Signal ratios $R$ and M1 matrix elements for the $5p_{3/2}(F'$=0$)\rightarrow 8p_{1/2}(F''$=1) and $5p_{3/2}(F'$=0$)\rightarrow 8p_{1/2}$\protect\linebreak[1]$(F''$=2) transitions.}

\begin{ruledtabular}
\begin{tabular}{l c c c}

 & Based on & Present & Present\\
 & Ref. \cite{Bayram2000} & Experiment & Theory\\
\hline
M1/E2 ratio, $R$  &  $0.38\pm0.03$ & $<0.007$ &0.0006 \\ \\
M1 matrix\\ element (10$^{-3}$ a.u.)  &  $55\pm1.4$ & $<7.4$ & 2.16 \\
 \end{tabular}
\end{ruledtabular}
\end{table}

\section{\label{sec:conclusions}Conclusions}

	We use high-resolution laser spectroscopy of ultracold $^{87}$Rb atoms to set an upper limit on the M1 contribution to the $5p-8p$ transition. Starting in an $F'$=0 state, we take advantage of the hyperfine selection rules, which allow for complete separation of the E2 and M1 processes. Our upper limit on the M1 transition strength is consistent with the theoretical prediction presented here, but disagrees with a previous measurement based on the linear polarization degree of the transition \cite{Bayram2000}.

	Our present limit is based on the signal for the M1-allowed hyperfine transition being $<$0.007 of that corresponding to the observed E2-allowed hyperfine transition. With an improved signal-to-noise ratio, the M1 transition should actually be observable, which would allow the theoretically predicted ratio of $R$=0.0006 to be tested. However, we are exciting the atoms to $8p_{1/2}$ by a two-photon process where the first step is resonant with the $5s_{1/2}(F$=1)$\rightarrow 5p_{3/2}(F'$=0) transition. When the second step is tuned to the M1-allowed $5p_{3/2}(F'$=0$)\rightarrow 8p_{1/2}(F''$=1) transition, the $5s_{1/2}(F$=1$)\rightarrow 5p_{3/2}(F'$=1$)\rightarrow 8p_{1/2}(F''$=1) two-photon transition, where the second step is E2-allowed, is also two-photon resonant, but with the $F'$=1 intermediate state off resonance by 72 MHz. Accounting for this detuning and the relative hyperfine transition strengths, we estimate that this off-resonant two-photon E2 contribution to the signal at the M1 resonance would give an apparent $R$=0.0007, which is coincidentally quite close to the predicted M1 contribution. These contributions could be distinguished by varying the detuning of the first step of the excitation. We note that Stark mixing of $s$ or $d$ character into the $8p$ level by an external field could give rise to a weak E1 contribution at the location of the M1 transition. However, this mixing is negligible under our conditions. An alternative method for measuring the M1 contribution would take advantage of the fine-structure selection rules for M1 and E2. The $5p_{1/2}\rightarrow 8p_{1/2}$ transition at 579.5 nm is M1 allowed but E2 forbidden, so a similar experiment to that presented here, but without the requirement of hyperfine resolution, could be performed. Light at 795.0 nm would be needed to populate $5p_{1/2}$.

	Our technique for measuring M1 matrix elements can be extended to other states in $^{87}$Rb and to other atomic systems possessing an $F'$=0 state. For example, the radioactive isotopes $^{223}$Fr and $^{225}$Fr both have nuclear spin $I$=3/2 and therefore an $F'$=0 level in their $7p_{3/2}$ state. Since relativistic effects increase with atomic number $Z$, heavier atoms such as francium will have larger M1 matrix elements, thus serving as good test cases for atomic structure calculations.

\begin{acknowledgments}
This work was supported in part by NSF grants PHY-0457126 and PHY-0653392. We acknowledge helpful discussions with Michael Moore regarding three-level system lineshapes.

\end {acknowledgments}

%\section*{References}

\end{document}